# Predicting a Ferrimagnetic Phase of $Zn_2FeOsO_6$ with Strong Magnetoelectric Coupling

P. S. Wang,[1] W. Ren,[2] L. Bellaiche,[3] and H. J. Xiang*[1]

[1]Key Laboratory of Computational Physical Sciences (Ministry of Education), State Key Laboratory of Surface Physics, Collaborative Innovation Center of Advanced Microstructures, and Department of Physics, Fudan University, Shanghai 200433, P. R. China

[2]Department of Physics, Shanghai University, 99 Shangda Road, Shanghai 200444, P. R. China

[3]Physics Department and Institute for Nanoscience and Engineering, University of Arkansas, Fayetteville, Arkansas 72701, USA

**Abstract**

Multiferroic materials, in which ferroelectric and magnetic ordering coexist, are of fundamental interest for the development of novel memory devices that allow for electrical writing and non-destructive magnetic readout operation. The great challenge is to create room temperature multiferroic materials with strongly coupled ferroelectric and ferromagnetic (or ferrimagnetic) orderings. $BiFeO_3$ has been the most heavily investigated single-phase multiferroic to date due to the coexistence of its magnetic order and ferroelectric order at room temperature. However, there is no net magnetic moment in the cycloidal (antiferromagnetic-like) magnetic state of bulk $BiFeO_3$, which severely limits its realistic applications in electric field controlled spintronic devices. Here, we predict that double perovskite $Zn_2FeOsO_6$ is a new multiferroic with properties superior to $BiFeO_3$. First, there are strong ferroelectricity and strong ferrimagnetism at room temperature in $Zn_2FeOsO_6$. Second, the easy-plane of the spontaneous magnetization can be switched by an external electric field, evidencing the strong magnetoelectric coupling existing in this system. Our results suggest that ferrimagnetic 3d-5d double perovskite may therefore be used to achieve voltage control of magnetism in future spintronic devices.

The highly efficient control of magnetism by an electric field in a solid may widen the bottle-neck of the state-of-the-art spin-electronics (spintronics) technology, such as magnetic storage and magnetic random-access memory. Multiferroics [1-7], which show simultaneous ferroelectric and magnetic orderings, provide an ideal platform for the electric field control of magnetism because of the coupling between their dual order parameters. For realistic applications, one needs to design/discover room temperature multiferroic materials with strong coupled ferroelectric and ferromagnetic (or ferrimagnetic) ordering.

Perovskite-structure bismuth ferrite ($BiFeO_3$) is currently the most studied room temperature single-phase multiferroic, mostly because its large polarization and high ferroelectric Curie temperature (~820 °C) make it appealing for applications in ferroelectric non-volatile memories and high temperature electronics. Bulk $BiFeO_3$ is an antiferromagnet with Néel temperature $T_N \approx 643$ K [8]. The Fe magnetic moments order almost in a checkerboard G-type manner with a cycloidal spiral spin structure in which the antiferromagnetic (AFM) axis rotates through the crystal with an incommensurate long-wavelength period [9]. This spiral spin structure leads to a cancellation of any macroscopic magnetization. The magnetic properties of $BiFeO_3$ thin films were found to be markedly different from those of the bulk: The spiral spin structure seems to be suppressed and a weak magnetization appears [10]. Nevertheless, the magnetization is too small for many applications [11]. In addition, an interesting low-field magnetoelectric (ME) effect at room temperature was discovered in Z-type hexaferrite $Sr_3Co_2F_{24}O_{41}$ by Kitagawa *et al.* [12]. Unfortunately, the electric polarization (about 20 $\mu C/m^2$) induced by the spin order is too low [13].

In searching for new multiferroic compounds, the double perovskite system was proposed as a promising candidate [14, 15]. The double perovskite structure $A_2BB'O_6$ is derived from the $ABO_3$ perovskite structure. The two cations B and B′ occupy the octahedral B sites of perovskite with the rock salt ordering. Double perovskite $Bi_2NiMnO_6$ was successfully synthesized under high-pressure, which displays the multiferroic behavior with a high ferroelectric transition temperature (485 K) but a low ferromagnetic transition temperature (140 K) [14]. Polar $LiNbO_3$ (LN)-type $Mn_2FeMO_6$ (M=Nb, Ta) compounds were prepared at 1573 K under 7 GPa [16]. Unfortunately, the magnetic ground state of $Mn_2FeMO_6$ is AFM with a rather low Néel temperature (around 80 K). Very recently, LN-type polar magnetic $Zn_2FeTaO_6$ was obtained via high pressure and temperature synthesis [17]. The AFM magnetic transition temperature ($T_N \sim 22$ K) for $Zn_2FeTaO_6$ is also low. In a pioneering work, Ležaić and Spaldin proposed to design multiferroics based on 3d-5d ordered double perovskites [18]. They found that $Bi_2NiReO_6$ and $Bi_2MnReO_6$ are insulating and exhibit a robust ferrimagnetism that persists above room temperature. Although coherent heteroepitaxy strain may stabilize the R3 ferroelectric (FE) state, free-standing bulk of $Bi_2NiReO_6$ and $Bi_2MnReO_6$ unfortunately take the non-polar P21/n structure as the ground state. The magnetic properties of non-polar double perovskites $Ca_2FeOsO_6$ [19] and $Sr_2FeOsO_6$ [20] were also theoretically investigated. Recently, Zhao *et al.* predicted that double perovskite superlattices $R_2NiMnO_6/La_2NiMnO_6$ (R is a rare-earth ion) exhibit an electrical polarization and strong ferromagnetic order near room temperature [21].

However, the ME coupling in these superlattices appears to be weak and the polarization to be small.

In this work, we predict that double perovskite $Zn_2FeOsO_6$ takes the FE LN-type structure as the ground state through a global structure searching. Similar to $Bi_2NiReO_6$ and $Bi_2MnReO_6$, $Zn_2FeOsO_6$ exhibit a strong ferrimagnetism at room temperature. Importantly, there is a rather strong magnetic anisotropy with the easy-plane of magnetization perpendicular to the FE polarization due to the presence of the significant 3d-5d Dzyaloshinskii-Moriya (DM) interaction. This suggests that the switching between the 71° or 109° FE domains by the electric field will cause the rotation of the magnetic easy-plane. Our work therefore indicates that $Zn_2FeOsO_6$ may be a material of choice for realizing voltage control of magnetism at room temperature.

It is well-known that there are several lattice instabilities including ferroelectric distortions and oxygen octahedron rotations in perovskite materials. We now examine how double-perovskite $Zn_2FeOsO_6$ distorts to lower the total energy. For this purpose, we perform a global search for the lowest energy structure based on the genetic algorithm (GA) specially designed for finding the optimal structural distortion [22]. We repeat the simulations three times. All three simulations consistently show that the polar rhombohedral structure with the R3 space group [shown in Fig. 1(b) and (c)] has the lowest energy for $Zn_2FeOsO_6$. Similar to $Zn_2FeTaO_6$, $Zn_2FeOsO_6$ with the R3 structure is based on the R3c LN-type structure. Previous experiments showed that double perovskite structure $A_2BB'O_6$ may adopt other structures, such as the P21/n [23], $R\bar{3}$ [24], and C2 structures [25]. Our density functional theory (DFT) calculations show that the R3 phase of $Zn_2FeOsO_6$ has a lower energy than the P21/n, $R\bar{3}$, and C2 structures by 0.22, 0.09, 0.45 eV/f.u., respectively. This strongly suggests that double perovskite $Zn_2FeOsO_6$ adopts the R3 structure as its ground state. This can be understood by using the tolerance factor defined for the LN-type $ABO_3$ system. It was shown that when the tolerance factor

$(t_R = \dfrac{r_A + r_O}{\sqrt{2}(r_B + r_O)}$, where $r_A$, $r_B$ and $r_O$ are the ionic radii of the A-site ion, B-site ion and O ion)

is smaller than 1, the polar $R3c$ structure is more stable than the $R\bar{3}c$ $ABO_3$ structure due to the A-site instability [26]. In the case of $Zn_2FeOsO_6$, the average tolerance factor (0.75) is smaller than 1, which suggests that the $R3$ structure is more stable than the non-polar $R\bar{3}$ structure. Note that phonon calculations shows that the $R3$ state of $Zn_2FeOsO_6$ is dynamically stable [27]. Additional tests indicate that the Fe and Os ions tend to order in a rocksalt manner (i.e., double perovskite configuration) to lower the Coulomb interaction energy [27,28].

Our electronic structure calculation shows that the R3 phase of $Zn_2FeOsO_6$ in the ferrimagnetic state is insulating [27]. The density of states plot shows that the Fe majority 3d states are almost fully occupied, while the minority states are almost empty. This suggests that the Fe ion takes the high-spin $Fe^{3+}$ ($d^5$) valence state. It is also clear that the Os ion takes the high-spin $Os^{5+}$ ($d^3$) valence state, which contrasts with the case of $Ba_2NaOsO_6$ where the Os atom takes a $5d^1$ valence electron configuration [29]. This is also consistent with the total magnetic moment of 2 $\mu_B$/f.u. for the ferrimagnetic state from the collinear spin-polarized calculation. Through the four-state mapping approach which is able to deal with spin interactions between two different atomic types [30], we compute the symmetric exchange parameters to find that the magnetic ground state of R3 $Zn_2FeOsO_6$ is indeed ferrimagnetic. The Fe-Os superexchange interactions mediated by the corner-sharing O ions [$J_1$ and $J_2$, see Fig. 1(c)] are strongly AFM ($J_1$ = 31.68 meV, $J_2$ = 29.62 meV). Here, the spin interaction parameters are effective by setting the spin values of $Fe^{3+}$ and $Os^{5+}$ to 1. Our results seem to be in contradiction with the Goodenough-Kanamori rule which predicts a ferromagnetic interaction between a $d^5$ ion and a $d^3$ ion since the virtual electron transfer from a half-filled σ-bond e orbital on the $d^5$ ion to an empty e orbital on the $d^3$ ion dominates the antiferromagnetic π-bonding t-electron transfer [31]. This discrepancy is because the <Fe-O-Os angle (about 140°) in $Zn_2FeOsO_6$ is much smaller than 180° (which is the case for which the Goodenough-Kanamori rule applies). The AFM super-superexchange interaction [$J_3$ and $J_4$, see Fig. 1(c)] between the Os ions is much weaker ($J_3$ =7.25 meV, $J_4$ =3.24 meV) than $J_1$ and $J_2$, but not negligible. Since the Fe 3d orbitals are much more localized than Os 5d orbitals, the super-superexchange interactions between the Fe ions are negligible. The dominance of the AFM Fe-Os exchange interactions suggests that the magnetic ground state of R3 $Zn_2FeOsO_6$ should be ferrimagnetic, which is also confirmed by the Monte Carlo (MC) simulations to be discussed below.

The magnetic anisotropy in R3 $Zn_2FeOsO_6$ is investigated by including spin-orbit coupling (SOC) in the DFT calculation. The DFT+U+SOC calculation shows that the magnetic moments tend to lie in the hexagonal ab-plane. Interestingly, when magnetic moments are in the ab-plane, there is a spin canting between the Fe 3d moments and Os 5d moments. We perform a test calculation to explicitly illustrate this point. We fix the Fe spin moments along the x-direction, and compute the total energies as a function of the angle (α) between Fe and Os moments. Fig. 2(a) shows that the total energy has a minima at $α_{min}$ ≈ 174° and the total energy curve is asymmetric with respect to the axis of α = 180°. This is solely due to the SOC effect since the total energy from the DFT+U calculations has a minima at α = 180°. The deviation of $α_{min}$ by 6° from 180° implies that the magnetic moment also acquires a weak component along the y-axis, which can be naturally understood by the universal law [32]. As shown in Fig. 2(b), the ferrimagnetic state with the moments along the c-axis is higher in energy by 0.55 meV/f.u. than the ferrimagnetic state with the Fe and Os moments aligned oppositely in the ab-plane ($\alpha = 180°$). The canting of the spins ($\alpha = 174°$) further lowers the total energy by 0.97 meV/f.u.. Therefore, the spin canting is extremely important for the easy-plane behavior.

To understand the magnetic anisotropy in R3 $Zn_2FeOsO_6$, we consider the general spin Hamiltonian [30]: $H_{spin} = \sum_{i<j} J_{ij} S_i \cdot S_j + \sum_{i<j} \vec{D}_{ij} \cdot (S_i \times S_j) + \sum_i A_i S_{iz}^2$, which includes symmetric exchange interactions, antisymmetric DM interactions, and single-ion anisotropy. For the magnetic structure in which the magnetic unit cell only contains one Fe ion and one Os ion, the total spin interaction energy can be written as:

$$E_{spin} = 3(J_1 + J_2) S_{Fe} \cdot S_{Os} + \sum_{i=1,6} D_{0i} \cdot (S_{Fe} \times S_{Os}) + A_{Fe}(S_{Fe}^z)^2 + A_{Os}(S_{Os}^z)^2,$$

where the DM interaction vectors $D_{0i}$ are shown in the insert of Fig. 2(a), $A_{Fe}$ and $A_{Os}$ are the effective single-ion anisotropy parameters. Here, the DM interactions between two different kinds of atoms are considered [33-35]. Our calculations show that $A_{Fe} = 0.89$ meV and $A_{Os} = -0.34$ meV, suggesting an overall weak easy-plane behavior resulting from the single-ion anisotropy term. The exchange interaction between the Os ions is omitted since it is a constant for such magnetic structure. Our analysis shows that the DM interaction results in the spin canting, and subsequently contributes dominantly to the easy-plane anisotropy [27].

By performing the parallel tempering (PT) MC simulation using the full spin Hamiltonian (containing symmetric exchange interactions, antisymmetric DM interactions, and single-ion anisotropy), we obtain the thermodynamic behavior of the magnetism in the R3 phase of Zn$_2$FeOsO$_6$ [36]. As shown in Fig. 3, there is a peak at 394 K in the specific heat curve, indicating that there is a magnetic phase transition at 394 K. The total spin moment (M$_{ab}$) in the ab-plane is also plotted as a function of temperature. We can see that the in-plane total spin moment increases rapidly near the magnetic phase transition point. This suggests that the in-plane total spin moment can be chosen as the order parameter, and the paramagnetic phase transforms to the ferrimagnetic phase with moments in the *ab*-plane at 394 K. We notice that the in-plane total spin moment is larger than 2 μ$_B$/f.u. at low temperature. This is a consequence of the spin canting resulting from the DM interactions. Our MC simulation suggests that R3 Zn$_2$FeOsO$_6$ is a room temperature ferrimagnet. Note also that test calculations show that the ferrimagnetic transition temperatures of Zn$_2$FeOsO$_6$ with the low-energy phases $R\bar{3}$ and P2$_1$/n are 344 K and 356 K, respectively. This indicates that the effect of the ferroelectric displacement, oxygen octahedron tilts, and the associated magnetoelectric coupling on the ferrimagnetic transition temperature is rather weak and does not prevent Zn$_2$FeOsO$_6$ from having a magnetic transition temperature above 300K.

Similar to the LiNbO$_3$ case where the paraelectric structure has the R$\bar{3}$c symmetry, the paraelectric state corresponding to the FE R3 Zn$_2$FeOsO$_6$ can be chosen to be $R\bar{3}$. Figure 4(a) shows the double well potential for Zn$_2$FeOsO$_6$. We can see that the barrier between the two FE state with opposite electric polarizations along the pseudo-cubic direction [111] is 0.09 eV/f.u., which is smaller than the value (about 0.20 eV/f.u.) in the case of PbTiO$_3$ [37]. Taking the $R\bar{3}$ structure as the reference structure, the electric polarization of R3 Zn$_2$FeOsO$_6$ is computed to be 54.7 μC/cm$^2$, which is much higher than that (25.0 μC/cm$^2$) in bulk tetragonal BaTiO$_3$ [38]. Thus, R3 Zn$_2$FeOsO$_6$ should be a switchable ferroelectric with a high electric polarization. Our tests based on an effective model Hamiltonian and molecular dynamics simulations show that the ferroelectric transition temperature of $Zn_2FeOsO_6$ is well above room temperature (see Part 9 and 10 of [27]).

Let us now discuss the possible ME coupling in double-perovskite $Zn_2FeOsO_6$. Because of the cubic symmetry of the perovskite structure, there are eight symmetrically equivalent <111> directions. In rhombohedral $Zn_2FeOsO_6$, the spontaneous electric polarization is directed along one of the eight <111> axes of the perovskite structure. Thus, in the sample of double-perovskite $Zn_2FeOsO_6$, there might occur eight different FE domains [see Fig. 4(b)]. Our above calculations show that the easy-plane of magnetization is always perpendicular to the direction of the electric polarization. Although a 180° switching of the ferroelectric polarization should not affect the magnetic state, a 71° or 109° switch of the FE domains by the electric field will change the orientation of the easy-plane of magnetization, as shown schematically in Fig. 4(c). This could be a promising route to manipulate the orientation of the ferrimagnetism by an electric field. A similar ME coupling mechanism in $BiFeO_3$ thin films has been demonstrated experimentally by Zhao *et al.*, who showed that the AFM plane can be switched by an electric field [39]. Note that magnetoelectric effects can be classified into two different types: one for which changing the magnitude of the polarization affects the magnitude of the magnetization (energy of the form $P^2 \cdot M^2$) and one for which changing the direction of $\vec{P}$ changes the direction of $\vec{M}$ (energy of the form $\vec{P} \times \vec{M}$). In $Zn_2FeOsO_6$, the first type of ME effect is weak, while the second type of ME effect is strong.

We now compare $Zn_2FeOsO_6$ with the classic multiferroic $BiFeO_3$. First, they adopt similar rhombohedral structures. Second, both compounds have high electrical polarizations. Third, both compounds are room temperature multiferroics. Fourth, the ME coupling mechanism is rather similar in that the magnetic easy-plane can be manipulated by electric field. However, the magnetic ground state of R3 $Zn_2FeOsO_6$ is dramatically different from $BiFeO_3$. $Zn_2FeOsO_6$ has a ferrimagnetic ground state, while $BiFeO_3$ is AFM. And the magnetic anisotropy in $Zn_2FeOsO_6$ is stronger than that (0.2 meV when U(Fe) = 5 eV [40]) in $BiFeO_3$ because of the strong SOC effect of the 5d Os ion. These desirable properties make $Zn_2FeOsO_6$ suitable for realizing electric-field control of magnetism at room temperature.

The reason why we practically proposed $Zn_2FeOsO_6$ as a possible multiferroic is two-fold. First, polar $Zn_2FeTaO_6$ has already been synthesized under high-pressure, as mentioned above. Second, it was experimentally showed that $Ca_2FeOsO_6$ crystallizes into an ordered double-perovskite structure with a space group of P21/n under high-pressure and high-temperature, and $Ca_2FeOsO_6$ presents a long-range ferrimagnetic transition above room temperature ($T_c$ ~320 K) [24]. Therefore, we expect that $Zn_2FeOsO_6$ is synthesizable and would most likely exhibit both ferroelectricity and ferrimagnetism at room temperature.

Work at Fudan was supported by NSFC, FANEDD, NCET-10-0351, Research Program of Shanghai Municipality and MOE, the Special Funds for Major State Basic Research, Program for Professor of Special Appointment (Eastern Scholar), and Fok Ying Tung Education Foundation. L.B. thanks the Department of Energy, Office of Basic Energy Sciences, under contract ER-46612.

e-mail: hxiang@fudan.edu.cn

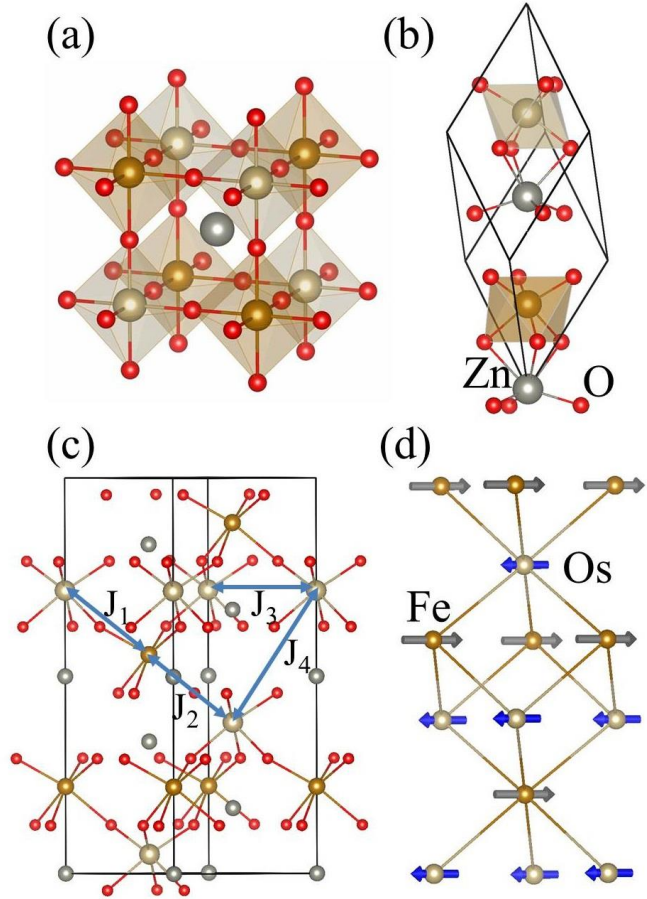

FIG. 1(color online). (a) The pseudo-cubic structure of double perovskite $Zn_2FeOsO_6$. (b) The polar rhombohedral structure of R3 $Zn_2FeOsO_6$. (c) The Fe-Os superexchange interactions mediated by the corner-sharing O ions ($J_1$ and $J_2$) and the AFM super-superexchange interaction ($J_3$ and $J_4$) between the Os ions. (d) The ferrimagnetic structure with the Fe and Os spin moments aligned in the ab-plane.

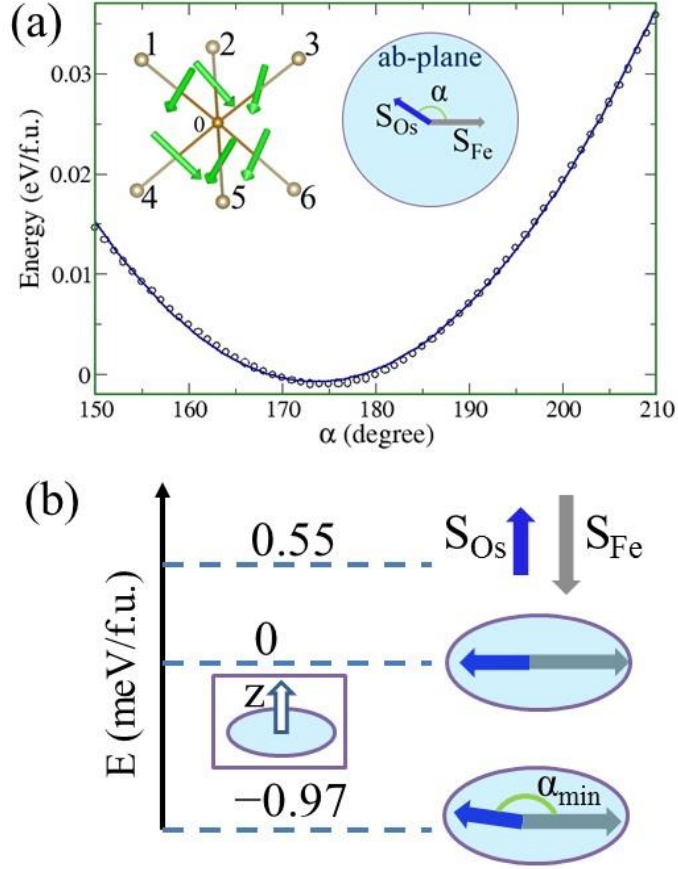

FIG. 2(color online). (a) The total energies as a function of the angle α [see the insert] from the direct DFT+U+SOC calculations (small circles). The total energy has a minima at $α_{min} ≈ 174°$. The total energy curve can be described rather well by the formula $E_{spin} = 3(J_1 + J_2)\cos α + 3(D^z_{01} + D^z_{04})\sin α$ (blue line). The insert shows the DM interaction vectors $\mathbf{D}_{0i}$ (i =1, 6) between a Fe ion and six Os ions and the definition of the angle α between the Fe spin and Os spin in the ab-plane. (b) Relative energies between different spin configurations. The ferrimagnetic state with the moments along the c-axis is higher in energy by 0.55 meV/f.u. than the ferrimagnetic state with the Fe and Os moments aligned oppositely in the ab-plane (α = 180°), and the canting of the spins (α = 174°) further lowers the total energy by 0.97 meV/f.u..

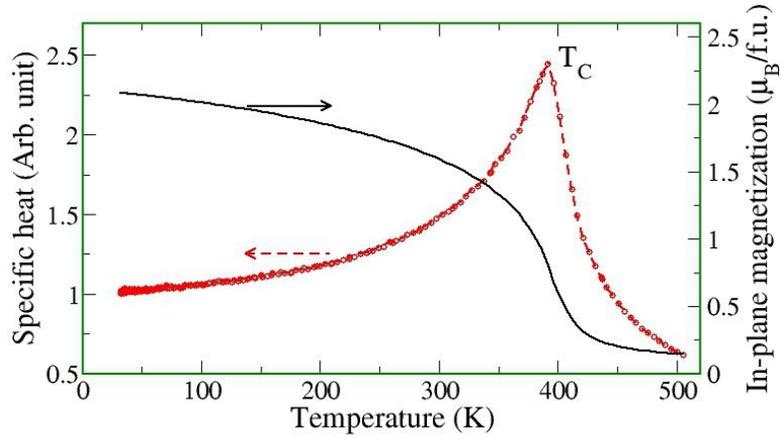

FIG. 3(color online). The specific heat and the total in-plane spin moment ($M_{ab}$) as a function of temperature from the PTMC simulations. The specific heat curve indicates that the ferrimagnetic Curie temperature ($T_c$) is 394 K. The in-plane total spin moment increases rapidly near $T_c$.

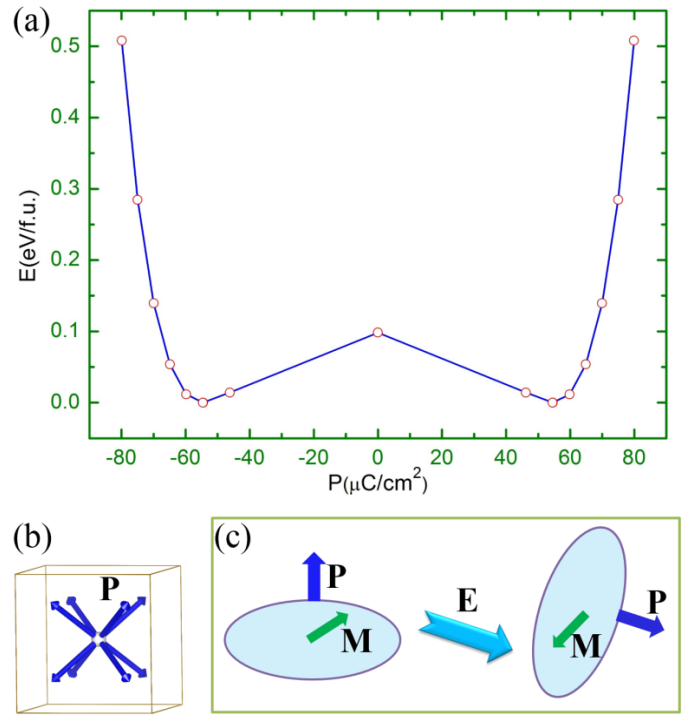

FIG. 4(color online). (a) The total energy as a function of the electric polarization for $Zn_2FeOsO_6$. It displays the double well potential with an energy barrier of 0.09 eV/f.u.. (b) Eight possible orientations of the FE polarization vector (**P**) in the sample of double-perovskite $Zn_2FeOsO_6$. (c) Illustration of the ME coupling in $Zn_2FeOsO_6$. A 71° or 109° switch of the FE domains by the external electric field will be associated with the reorientation of the easy-plane of magnetization.